\documentclass[sigplan,nonacm]{acmart}
\usepackage{amssymb}

\usepackage[capitalise, nameinlink]{cleveref}
\usepackage{stfloats}
\usepackage{pgfplots}
\usepackage{subcaption}
\usepackage{xcolor}
\pgfplotsset{compat=1.18}

\definecolor{darkgreen}{RGB}{0,128,0}
\definecolor{darkred}{RGB}{178,34,34}
\definecolor{lightorange}{RGB}{255,204,102}

\usepackage{tikz}
\usepackage{wrapfig}
\usepackage{graphicx}
\usepackage[normalem]{ulem}
\usepackage{algorithm}
\usepackage{algpseudocode}
\usepackage[]{markdown}

\newcommand{\beginbsec}[1]{\noindent\textbf{#1. \hspace{2pt}}}

\newcommand{\squishlistnum}{  
 \newcounter{qcounter}
 \begin{list}{\roman{qcounter})~}{\usecounter{qcounter}}
  { \setlength{\itemsep}{0pt}
     \setlength{\parsep}{0pt}
     \setlength{\topsep}{0pt}
     \setlength{\partopsep}{0pt}
     \setlength{\leftmargin}{0em}
     \setlength{\labelwidth}{0em}
     \setlength{\labelsep}{0em} } }

\newcommand{\squishlist}{ 
 \begin{list}{$\bullet$}
  { \setlength{\itemsep}{3pt}
     \setlength{\parsep}{0pt}
     \setlength{\topsep}{3pt}
     \setlength{\partopsep}{0pt}
     \setlength{\leftmargin}{1em}
     \setlength{\labelwidth}{1em}
     \setlength{\labelsep}{0.5em} } }

\newcommand{\squishend}{
  \end{list}  }

\ifdefined\RELEASE
    \newcommand\old[1]{}
    \newcommand\boris[1]{}
    \newcommand\david[1]{}
    \newcommand\dilina[1]{}
    \newcommand\shyam[1]{}
    \newcommand\shengda[1]{}
    \newcommand\amir[1]{}
    \newcommand\antonio[1]{}
    \newcommand\todo[1]{}
    
    \newcommand\invisible[1]{}

\else
    \newcommand\old[1]{{\color{gray}[OLD]: #1}}
    \newcommand\boris[1]{{\color{blue}[Boris]: #1}}
    \newcommand\david[1]{{\color{orange}[David]: #1}}
    \newcommand\dilina[1]{{\color{purple}[Dilina]: #1}}
    \newcommand\shyam[1]{{\color{teal}[Shyam]: #1}}
    \newcommand\shengda[1]{{\color{olive}[Shengda]: #1}}

    \newcommand\amir[1]{{\color{magenta}[Amir]: #1}}
    \newcommand\antonio[1]{{\color{cyan}[Antonio]: #1}}
    \newcommand\todo[1]{{\color{orange}[TODO]: #1}}
    
    \newcommand\invisible[1]{}
\fi

\newcommand{\coolName}{Odyssey}
\newcommand{\optimizer}{Intelligent Plan Explorer}
\newcommand{\optimizerAcronym}{IPE}

\newcommand{\stockPlanner}{Stock Planner}
\newcommand{\logicalPlan}{Operator Ordering Plan}

\begin{document}

\title{Odyssey: An End-to-End System for Pareto-Optimal Serverless Query Processing}

\author{Shengda Zhu}
\authornote{Both authors contributed equally to this research.}
\author{Shyam Jesalpura}
\authornotemark[1]
\affiliation{%
  \institution{The University of Edinburgh}
  \city{Edinburgh}
  \country{UK}
}

\author{Amir Shaikhha}
\affiliation{%
  \institution{The University of Edinburgh}
  \city{Edinburgh}
  \country{UK}}

\author{Antonio Barbalace}
\affiliation{%
  \institution{The University of Edinburgh}
  \city{Edinburgh}
  \country{UK}
}

\author{Boris Grot}
\affiliation{%
 \institution{The University of Edinburgh}
 \city{Edinburgh}
 \country{UK}}

\begin{abstract}

Running data analytics queries on serverless (FaaS) workers has been shown to be cost- and performance-efficient for a variety of real-world scenarios, including intermittent query arrival patterns and sudden load spikes that overwhelm a deployed cluster of VMs. Alas, existing serverless data analytics
works focus primarily on the serverless execution engine and naively assume that a "good" query execution plan somehow exists. Meanwhile, even simple analytics queries on serverless have a huge space of possible plans, with vast differences in both performance and cost among plans.

This paper introduces \coolName{}, an end-to-end serverless-native data analytics pipeline that integrates a query planner, cost model and execution engine. 
\coolName{}~automatically generates and evaluates serverless query plans, utilizing state space pruning heuristics and a  novel search algorithm to identify Pareto-optimal plans that balance cost and performance with low latency even for complex queries.
Our evaluations demonstrate that \coolName{} accurately predicts both monetary cost and latency, and consistently outperforms AWS Athena on cost and/or latency. 

\end{abstract}
\pagestyle{plain}
\maketitle

\section{Introduction}
\label{sec:intro}
\raggedbottom

The cloud data analytics market is booming due to increasing data volume and demand for online processing~\cite{analytics-market}. Running data analytics on VM clusters is cost- and performance-effective under predictable load, but is cost inefficient under transient load and has poor performance under sudden load spikes.
Query-as-a-Service (QaaS) solutions like Amazon Athena~\cite{athena} or Google BigQuery~\cite{bigquery} seemingly address these concerns. 
These black-box engines take a query and a dataset as an input and output results along with a bill. 
However, QaaS  fails to provide either cost- or performance-predictability, making it impossible for users to reason about either.

Recent work has identified an opportunity in overcoming the limitations of existing
VM-based and QaaS offerings for executing data analytics queries using the {\em serverless}
 (i.e., FaaS) paradigm~\cite{lambada, starling}.
In the serverless model, jobs are deployed on light-weight stateless workers that are launched on-demand by the cloud provider.
The cloud provider is fully responsible for scaling worker instances based on real-time load, while users are billed only for the CPU time and memory usage of the workers running their code. This
combination of features is highly desired by users who value deployment simplicity, extreme elasticity and pay-per-use billing of serverless. Alas, today's serverless frameworks are not without limitations, which include a lack of direct function-to-function communication (requiring the use of intermediate
storage) as well as the high cost of serverless workers per unit of runtime as compared to VMs~\cite{higher-cost}.

Recent works have demonstrated that data analytics using serverless workers offers a viable alternative to VMs and QaaS engines for scenarios that include intermittent query arrivals or when a load spike overwhelms a VM cluster
~\cite{lambada, starling, cackle, pixel}. 
However, the focus of prior work has been on the serverless {\em execution engine}, which needs to launch the serverless workers upon query arrival, optimize data shuffle operations for the stateless workers, and deal with stragglers. 
In doing so, they assume that a serverless query plan is already provided to the engine. Such a plan defines the sequence of query operators, referred to as stages in this work (e.g., {\em scans} for reading data, {\em joins} for combining tables), along with the resources allocated to each stage.
Notably, the serverless query plans in these papers were manually produced or manually adapted from a conventional query plan with no indication as to their cost- or performance efficiency. 

We find that
the serverless environment offers a vast landscape of potential query plans even for simple queries, with massive cost and performance differences across various plans. 
These differences are shaped by parameters that include the number of serverless workers, the capabilities of each worker, the number of partitions for intermediate data, and the implementation of individual operators.
Moreover, in contrast to traditional DBMS systems (where a fixed set of resources is typically provisioned for all of the stages of a given query) the elastic nature of serverless presents an opportunity to tailor the resources for each individual query stage. 
Such per-stage specialization further increases the space of possible query plans. 
For instance, in AWS Lambda, on TPC-H Query 4 with a 1TB dataset, there are over a million potential query plans; the difference between the fastest and the slowest query plan is over 50x, while the cost difference between the most- and least-expensive is over 1000x. 

A key challenge in serverless analytics is not just identifying an optimal plan from a cost and performance perspective, but doing so with low latency since planning time is part of the end-to-end latency experienced by the user. This challenge is exacerbated by the need to process never-seen-before queries on arbitrary-sized datasets, rendering a-priori profiling as suggested by recent work~\cite{ditto} impractical. An additional complication in assessing potential query plans stems from the fact that today's FaaS infrastructure carries significant performance variability stemming from cold starts and storage contention; if not accounted for at query planning time, the efficiency of a selected plan can deviate from the estimate by a significant amount and render the selected plan sub-optimal. To the best of our knowledge, none of these issues have been addressed by prior work.

We introduce the Serverless qUery
Planner and ExecutoR (\coolName{}), an end-to-end data analytics pipeline designed specifically for serverless environments. 
It combines a query planner, cost model, and execution engine to efficiently process complex SQL queries.
When a query arrives, \coolName{}'s planner explores the space of potential query plans, considering  aspects such as the number of workers per stage, worker size, and storage type. To estimate the cost and latency of considered plans, the planner consults a query-agnostic cost model, which accounts for runtime variability due to cold starts and storage contention. Once a plan from the cost/performance Pareto frontier produced by the planner is chosen, it is executed by \coolName{}'s execution engine, which maximizes performance by overlapping compilation of later stages with interpreted execution of earlier stages.

In order to efficiently navigate the vast plan space, we introduce a set of serverless-specific heuristics that significantly reduce the number of configurations that need to be considered per each query stage. The remaining configurations are then fed to a novel algorithm that traverses the query stage by stage, 
eliminating all but the Pareto-optimal plans up to that stage. The algorithm reduces the state space of considered plans from exponential in the number of stages to approximately a constant, thereby enabling optimal plan selection with low latency and high memory efficiency.

We evaluate \coolName{} on a set of TPC-H queries with dataset sizes of up to 10TB. We find that \coolName{} can predict Pareto-optimal points on the cost-performance curve, and that these predictions are accurate when compared to actual executions. 
The time required by the planner to enumerate and prune the search space is under 5\% of the total execution time and is on-par with AWS Athena, a state-of-practice QaaS. Moreover, the preferred configurations identified by \coolName{} are superior to AWS Athena in terms of cost and/or performance.

\noindent
To summarize, this paper demonstrates the following:
\squishlist 

\item Serverless presents a vast optimization space of query plans with huge cost-performance trade-offs.
\item \coolName{} can automatically construct and evaluate the space of serverless plans for a SQL query, identify the pareto-optimal plans, and execute them. To perform these steps efficiently, \coolName{} deploys (1) a novel state pruning approach for rapid identification of cost/performance Pareto-optimal plans; (2) a query-agnostic cost model that takes into account performance variability inherent in the serverless environment; and (3) a hybrid execution strategy that hides compilation latency of operators for later stages with interpreted execution of earlier stages.

\item Evaluation on TPC-H queries with datasets up to 10TB reveals that \coolName{} can quickly and accurately predict Pareto-optimal configurations that consistently outperform AWS Athena in terms of cost, performance or both.
\squishend

\noindent
To encourage further research and reproducibility, we plan to open source \coolName{} upon publication.

\section{Motivation}
\label{sec:motivation}
\subsection{Query processing basics}
\label{sec:motivation:basics}

  
  
\begin{figure}[t]
  \centering
  \includegraphics[width=0.4\textwidth,trim=0 2.57cm 0 1.42cm,clip]{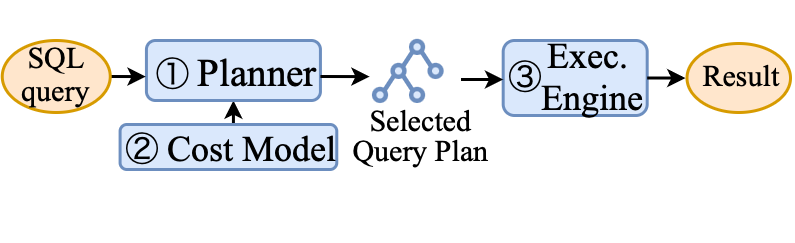}
  
  \caption{Basic end-to-end query processing flow.}
  
  \label{fig:query_plan_flow}
\end{figure}

Modern data analytics rely on a query execution framework to parse the incoming query, choose the preferred plan from a set of potential plans, execute it and return the results to the user.
This process, shown in \cref{fig:query_plan_flow}, consists of three major components:
(1) the {\em query planner} takes an input SQL query and data statistics to generate a search space of semantically equivalent plans. 
From this space, it selects one plan, typically optimizing for latency, cost, or a joint cost-performance trade-off;
(2) for each plan considered by the planner, the {\em cost model} estimates the execution time and monetary cost based on the available resources including the number of workers, CPU, memory, network bandwidth, etc.
It enables the planner to reason about cost-performance trade-offs across candidate plans;
(3) the {\em execution engine} executes the chosen query plan on available workers.
It implements the {\em query operators} (e.g., {\em scans} for reading data, {\em joins} for combining tables) while orchestrating workers, managing data movement, and producing the final results in the correct order.

\subsection{Query processing in the cloud}
\label{sec:motivation:cloud}

Cloud query processing has evolved due to growing data volumes and fluctuating query loads, pushing businesses toward 
analytics frameworks that run on VM clusters \cite{snowflake,redshift}. While clusters of VMs provide predictable performance when properly sized, they struggle in several common scenarios. Unexpected load spikes overwhelm clusters as autoscaling delays can exceed a minute, degrading user experience. Fluctuating workloads with burst-idle patterns reduce cost efficiency since VMs are billed regardless of utilization. Sporadic query needs, such as occasional analysis by scientists or small businesses, make VM cluster deployment difficult to justify given management overhead and expertise requirements. 

Query-as-a-Service (QaaS) offerings like AWS Athena and Google BigQuery present black-box alternative for these situations, allowing users to execute SQL queries on hosted data. However, the opaque nature of QaaS creates uncertainty about performance and cost expectations~\cite{starling}, potentially failing to meet users' needs on either front.

\begin{figure}[t]
    \centering
        \includegraphics[width=0.35\textwidth, trim=0cm 1.9cm 0cm 2.2cm,clip]{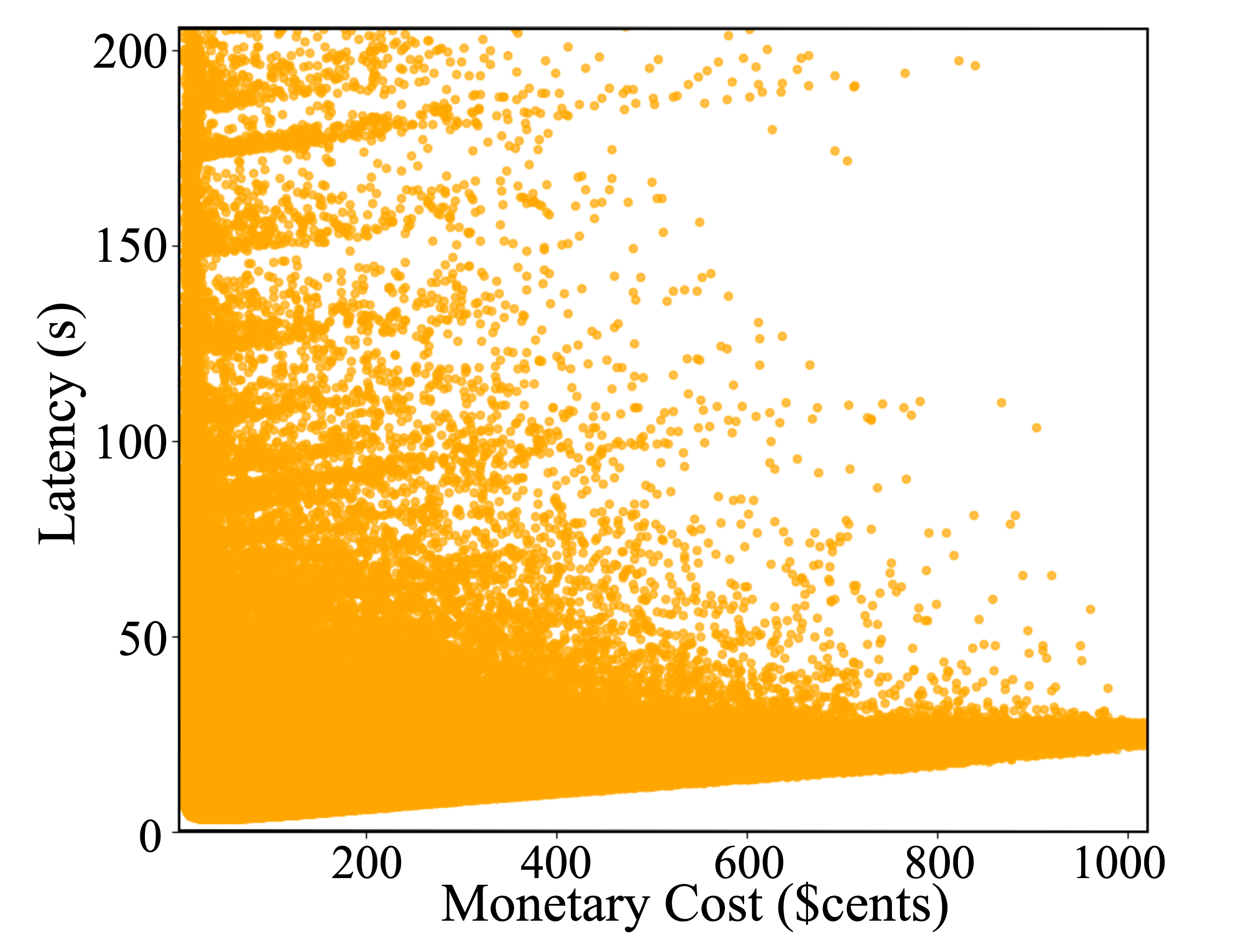}
    \caption{Cost and performance spectrum of query plans for TPC-H Q4 on 1TB dataset.}
    \label{fig:cost_spectrum}
\end{figure}

\subsection{Serverless to the rescue?}
\label{sec:motivation:prior_works}

The inherent limitations of VM-based clusters and opaque QaaS offerings emphasize the need for a query processing framework that can rapidly scale resource in response to a shift in demand while providing ease of management and cost efficiency. Prior work has shown that serverless computing can fit this bill with competitive cost and performance for intermittent query processing loads
\footnote{
  Effectiveness of serverless is not limited to intermittent queries, but current cloud pricing models make it unattractive for continuous query execution.} 
~\cite{pixel, lambada,starling,cackle}. 

Serverless computing has several strengths that makes it attractive for data analytics. 
Serverless workers are highly scalable and exhibit very low startup times on the order of seconds rather than a minute or more for VMs~\cite{cackle,ec2-boot,firecracker} -- a combination that makes serverless an extremely elastic compute resource and perfect for low-latency intermittent processing. Additionally, serverless workers are billed per-invocation according to the amount of CPU time and memory used, which eliminates expenses at idle periods. 

Alas, these advantages do not come without limitations. Serverless workers are stateless in nature and come with limited per-worker CPU and memory capacity (e.g., in AWS Lambda~\cite{lambda-cpu}, 1-6 cores and 128MB-10GB of memory).


Prior works on using today's serverless offerings for complex data analytics have focused on the serverless execution engines and/or task schedulers
, which can execute a given query plan in a way that exploits the strengths of the serverless workers while working around their limitations~\cite{lambada, starling, nimble}.
Notably, these works lack a query planner and a cost model to enable automated generation and assessment of potential query plans. Instead, a  "good" query plan was assumed to exist for a given query, which in practice means using hand-assembled or hand-tuned plans specifying the number of workers and other parameters -- a strategy that is not viable in a meaningful deployment. 

A recent paper has demonstrated a more complete system that includes a cost model that helps determine the preferred number of workers per query stage~\cite{ditto}. However, it relies on  query-specific profiling along with manually-specified parameters. These aspects mandate a-priori knowledge and profiling of queries and rely on the user to find parameters that are critical to cost/performance; both aspects simply not viable in a practical cloud deployment. 




\section{Challenges in Serverless Query Pipeline}
\label{sec:challenges}



Planning and executing complex SQL queries 
efficiently for a serverless environment without query-specific profiling or manual tuning is challenging.
These challenges differ significantly from traditional VM-based systems and span from query planning to reliable cost prediction in a noisy environment. We next discuss these challenges in detail.


\subsection{Gigantic search space}
\label{sec:challenges:search_space}
A primary challenge in serverless query processing lies in navigating the vast and complex optimization space to create an optimal query plan. 
Since the planning time is part of the total execution time experienced by the user; the space exploration must be completed with latency that is a small fraction of the execution time.


To illustrate the vast optimization space inherent in a serverless environment,
\cref{fig:cost_spectrum} visualizes a sampled space of serverless query plans for a single stage join query TPC-H Q4 (on 1TB dataset) on AWS Lambda with S3 storage. The figure shows that the query's cost (on the X-axis) and latency (on the Y-axis) can vary significantly depending on the query plan. 
After judiciously discretizing the search space (e.g., only considering meaningful memory sizes for each worker, sensible numbers of workers for each query stage etc), the total number of configurations exceeds a million with over 1000x difference in cost and over 50x in latency. The cost and performance difference 
in serverless query plans is driven by a multitude of parameters.
The extreme elasticity of serverless allows scaling from one to thousands of workers, each with configurable CPU and memory.
The need for storage between query stages further expands the space, with choices like storage types and number of partitions of intermediate data. 
To fully exploit the flexibility offered by serverless, these decisions should be made per query stage, causing the number of query plans to grow exponentially with the number of stages in a query.
For complex queries, we observe that merely enumerating all of the possible query plans takes minutes and can exceed server memory capacity \cref{sec:eval:optimizer:space}, as possible plans can easily surpass a billion.

\subsection{Dealing with arbitrary queries}
\label{sec:challenges:unknown}
We identify two major challenges in efficient planning and execution of queries that are unknown beforehand. The first challenge is estimating latency and cost of query plans for unseen queries and the second is achieving high-performance execution without hand-optimized code.

\noindent
\textbf{Planning for arbitrary queries.}
\label{sec:challenges:planning}
To navigate the vast search space of possible plans, an accurate cost model is required. Prior work in the serverless analytics space relied on
 query- and dataset-specific cost models by profiling individual queries~\cite{ditto}. Such an approach requires re-profiling and rebuilding the model for every new query and dataset size~\cite{ditto}. 
Designing a {\em query-agnostic} cost model for serverless substrate is a challenge given the variety and complexity of potential queries.

\noindent
\textbf{Executing arbitrary queries.}
\label{sec:challenges:engine}
Specializing the source code to use query-specific optimizations can unlock significant performance gains \cite{vectorwise,hyperdbcompiler}.
Prior works \cite{lambada,starling} 
relied on hand-written query plans for serverless query execution; however, this approach is not viable in practice.
In query processing systems, two main techniques have emerged for high-performance query execution engines: interpreted execution~\cite{photon,vectorwise} and just-in-time (JIT) compiled execution~\cite{hyperdbcompiler,querycompiler}. The two approaches have different trade-offs depending on the query~\cite{tpch-representative}. Short-running queries prefer interpreted execution to elide compilation time; long-running queries amortize compilation over query execution.
Choosing one of these approaches over the other is a challenge as the relative performance depends on the query and data characteristics, which are only known at runtime.

\subsection{Planning for the variabilities in execution}
\label{sec:challenges:cost_model}
To navigate the massive search space, a cost model is required to estimate the performance and monetary cost of a query plan without actually executing it. 
Creating such a cost model for serverless analytics is challenging because queries are unknown beforehand and the environment has unique performance dynamics.
Unlike traditional VM-based systems, serverless workers are ephemeral and lack direct function-to-function communication. 
These characteristics introduce variability in end-to-end execution that mush be accounted for by the cost model for accurate prediction. Two specific sources of variability are dominant in serverless:


\noindent
\textbf{Cold starts:} A serverless worker may experience a "cold start" if a new instance needs to be created. Even when workers are reused along the query processing pipeline, 
ignoring cold starts in the cost model causes query latency predictions to deviate from reality by more than 35\%.

\noindent
\textbf{Storage stragglers:} 
Due to the lack of direct function-to-function communication, serverless workers must 
communicate through a storage service that 
comes with a high degree of performance unpredictability due to contention and stragglers~\cite{s3-throttle}.
While runtime techniques such as retries can help, they cannot fully eliminate storage-layer variability.
When omitted from the cost model, storage stragglers cause query latency predictions to deviate from reality by over 40\%.

Prior works have proposed runtime techniques to mitigate cold starts and storage stragglers\cite{lambada, starling}, but we find these do not fully eliminate variability.
Ignoring these variabilities at planning time leads to execution plans that are over 2x more costly than when using a variability-aware cost model.



\begin{figure}[t]
    \includegraphics[width=0.43\textwidth,trim=0cm 0.65cm 0cm 0.32cm,clip]{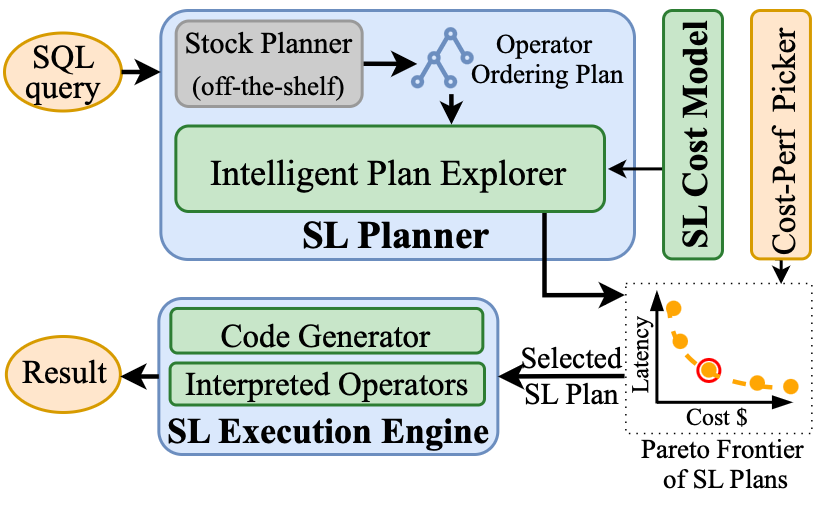}
    \caption{Overall design of \coolName{}.}
    \label{fig:design}
\end{figure}

\section{Design Overview}
\label{sec:design}

We introduce \coolName{}, an end-to-end system for serverless query planning and execution that tackles the challenges presented in~\cref{sec:challenges}. The system consists of three main components: the \textit{Serverless (SL) Planner}, the \textit{SL Cost Model}, and the \textit{SL Execution Engine} as illustrated in \cref{fig:design}.

\noindent
\textbf{SL Planner} is designed to identify Pareto-optimal plans for serverless query execution, balancing cost and performance. 
As it operates on the query execution critical path, it must produce SL execution plans quickly. 
To solve the gigantic search space challenge (\cref{sec:challenges:search_space}), we introduce \textit{\optimizer{} (\optimizerAcronym{})} which combines serverless-aware heuristics for reducing the search space with a novel algorithm for rapid search space exploration (\cref{sec:design:planner}).

\noindent
\textbf{SL Cost Model} is a query-agnostic function (\cref{sec:challenges:unknown}) that estimates the \textit{execution time} and \textit{monetary cost} of each plan variant without actual execution.
A key challenge is capturing runtime variability—especially cold starts and storage-layer effects (\cref{sec:challenges:cost_model})—which strongly impact both performance and cost (\cref{sec:design:costmodel}).

\noindent
\textbf{SL Execution Engine} is responsible for running the plan
chosen by the SL Planner, which specifies operators, worker
counts, CPU core allocations, and intermediate storage types.
The engine employs a hybrid strategy that combines interpreted and compiled execution (\cref{sec:challenges:unknown}), hiding compilation latency behind interpreted early stages to achieve near-compiled performance without start-up delays.



The complete \coolName{} pipeline leverages the above components to process analytics queries as follows.
The \textit{\stockPlanner{}} first takes an input query and generates an \texttt{\logicalPlan{}}. 
This logical plan is then passed to the \textit{\optimizerAcronym{}}, which prunes and explores the space of candidate SL execution plans.
For each plan variant, the \textit{SL Cost Model} estimates execution time and monetary cost, enabling the \textit{\optimizerAcronym{}} to identify Pareto-optimal \texttt{SL Plans}.
From the Pareto-optimal set, the \textit{\optimizerAcronym{}} automatically selects the plan that best aligns the user's cost-performance preferences.
Finally, the \textit{SL Execution Engine} receives the chosen \texttt{SL Plan}, generates the necessary code, and executes the query using a combination of interpreted operators and JIT-compiled operators.

\section{\coolName{} Details}


\subsection{SL Planner}
\label{sec:design:planner}

The \textit{SL Planner} takes an input query and produces Pareto-Optimal frontier of SL execution plans.
We first obtain the logical ordering of query operators using a standard stock planner (HyperDB~\cite{hyperdbcompiler}).
This stock planner estimates cardinality (input data sizes) for each stage from a representative data sample.
Our contribution lies in the next step: given this logical plan, \optimizer{} explores candidate SL execution plans by varying parameters such as worker count per stage.
For each candidate plan, the SL Cost Model estimates its execution time and monetary cost, effectively turning plan selection into an optimization problem: identifying the set of plans with the best cost/performance trade-off.

We first formalize this optimization problem, then present key insights that motivate a novel algorithm for efficiently producing the Pareto frontier.
To further reduce search overhead, we add serverless-aware heuristics that prune the search space at each query stage and integrate them into the algorithm to efficiently discover the Pareto frontier.



\subsubsection{The Challenge: Exponential Configuration Space}
\label{sec:design:optimizer:challenge}


Consider a query divided into $n$ stages. The planner considers the following set of parameters, for each stage $i \in [1,n]$: 

\squishlist
\item Number of workers ($\mathbf{w}_i$)
\item Worker size (memory capacity and number of cores) ($\mathbf{m}_i$)
\item Number of partitions used to split workers' output ($\mathbf{p}_i$)
\item Storage service type for intermediate data ($\mathbf{s}_i$)
\squishend

Let $|\mathcal{V}_i| = |\mathbf{w}_i \times \mathbf{m}_i \times \mathbf{p}_i \times \mathbf{s}_i|$ be the number of configurations for stage $i$. A complete query configuration requires selecting options for all stages, resulting in a total of $|\mathbf{\Omega}| = \prod_{i=1}^{n} |\mathcal{V}_i|$ configurations. This product grows exponentially with the number of stages ($n$), making an exhaustive search (evaluating each combination using the cost model from \cref{sec:design:costmodel}) computationally infeasible for complex queries.

The optimization problem is thus formally stated as: {\em Given the exponentially-large space $\mathbf{\Omega}$ of possible configurations for a given query, find the set of configurations $\mathcal{P}(\mathbf{\Omega})$ that lie on the cost-performance Pareto frontier}.

\subsubsection{Controlling Space Complexity}
\label{sec:design:optimizer:insight}
\noindent
This exponential growth is caused by inter-stage dependencies-- a preferred configuration for one stage depends on choices made in all of the previous stages. 
We address this challenge by leveraging two key insights that constrain these dependencies.

\squishlist
\item \textbf{Insight 1: Worker size has no bearing on configurations of other stages}:
The size of a producer worker (i.e., number of cores and memory allocation) does not influence the output it produces, nor the number of partitions into which the output is split. 
Consequently, worker size in a given stage has no effect on preferred configurations in the next stage, or further downstream, making worker size a {\em stage-confined effect}.

\item \textbf{Insight 2: Worker count only directly affects configurations of the adjacent stages}: 
The worker count in stage $i$ only directly affects its neighbor stages: stage $i-1$, which must produce its outputs, and stage $i+1$, which consumes its outputs.
This property makes worker count a {\em neighbor-confined effect}. 
Similarly, the choice of intermediate storage is neighbor-confined: the type of storage used by a producer stage $i$ impacts only the cost and performance of that stage and its immediate consumer stage.

\squishend

\noindent
These insights enable us to decompose the global optimization problem into manageable sub-problems at the granularity of individual stages (via stage-confined effects) and adjacent stage pairs (via neighbor-confined effects), thus dramatically reducing the search space and enabling fast Pareto frontier generation without sacrificing global optimality.

\subsubsection{State Space Pruning Heuristics}
\label{sec:design:optimizer:heuristics}
Building on this decomposition, we first apply the following heuristics to prune the search space for each query stage:

\squishlist 
\item \textbf{H1: Cardinality constraints:} Each stage enforces minimum and maximum input data sizes per worker to avoid excessive parallelism (leading to under-utilized workers) and prevent worker memory overflow (due to oversized inputs).
The cardinality constraints thus bound the number of workers per stage to $[w_{min},w_{max}]$.

\item \textbf{H2: Exponentially sampled worker count:} Within the range from H1, candidate worker counts are sampled exponentially:$[w_{min},w_{min}+2^1,w_{min}+2^2,\dots]$ up to $w_{max}$.
This effectively reduces the search space with negligible impact on identifying preferred configurations.

\item \textbf{H3: Integral core allocation:} Each worker is allocated an integer number of CPU cores; fractional cores cause unbalanced data distribution and idle resources. 
In AWS Lambda, the memory size requested determines the number of cores allocated \cite{lambda-cpu}, with each core corresponding to 1770 MB of memory.

\item \textbf{H4: Maximizing computing utilization:} When distributing the input dataset or intermediate data to workers, we assign data so that number of partitions per worker is a multiple of its number of cores. This eliminates idle cores and improves resource usage.
\item \textbf{H5: Partition alignment:} 
Matches the number of data partitions in stage $i$ to the worker count in stage $i+1$, ensuring even data distribution. 
If the partition count is lower or higher than the next stage's worker count, differences in the amount of intermediate data produced per worker can create stragglers, degrading query latency. 

\squishend

\noindent
With these heuristics, \coolName{} incrementally builds the feasible configuration space for each stage. \cref{alg:generate-stage-space} presents the pseudocode for this stage-space generation, forming the foundation for our full optimization algorithm.

\begin{algorithm} [t]
\caption{Stage Configuration Space Generation}
\label{alg:generate-stage-space}
\begin{algorithmic}[1]
    \footnotesize
    \State \textbf{Input:} Card - estimated input data size of stage $i$
    \State \textbf{Output:} $V$ - configuration space for stage $i$
\Procedure{GenStageSpace}{Card}
    \State $W \gets \text{ApplyHeuristics}(H_1, H_2, \text{Card})$ \Comment{$W$: worker counts}
    \State $M \gets \text{ApplyHeuristics}(H_3, H_4)$ \Comment{$M$: worker sizes}
    \State $S \gets \{\text{S3\_Standard}, \text{S3\_Onezone}, \ldots\}$ \Comment{$S$: storage type}
    \State $V \gets \{(w_i, s_i) : \boldsymbol{m}_i \mid w_i \in W, s_i \in S, \boldsymbol{m}_i \subset M\}$ 
    \Comment{Map: $\boldsymbol{m}_i$ grouped by $(w_i, s_i)$ key, denotes all valid worker sizes}
    \State $V.\text{constraint} \gets \text{``}p_i = w_{i+1}\text{''}$ \Comment{Partition unset, $H_5$}
    \State \Return $V$
\EndProcedure
\end{algorithmic}
\end{algorithm}

\subsubsection{Incremental Pareto Boundary Search} 
\label{sec:design:optimizer:algorithm}
While heuristics reduce the search space, they cannot fully address its exponential growth, which is a crucial issue for complex many-stage queries. 
To address this issue, we introduce the {\em Incremental Pareto Boundary Search algorithm}, which constructs the global cost-performance Pareto frontier by traversing query stages sequentially, generating locally Pareto-optimal configurations as it proceeds. At each stage, the algorithm takes the locally Pareto-optimal configurations carried over from the previous stage, extends them with current-stage parameters (e.g., worker counts after applying heuristics), and prunes configurations that cannot be part of the global frontier.
This incremental pruning transforms the exponential search space into one with size bounded by the local frontier, independent of the number of stages. This enables both space- and time-efficient global Pareto frontier generation.

\cref{alg:incremental-pruning} outlines the full algorithm. 
Line 8 ensures that configurations with neighbor-confined effects maintain their own local Pareto frontier until passed to the next stage, while line 16 performs the local pruning.

\begin{figure}[t]
    \includegraphics[width=0.45\textwidth,trim=0 0.8cm 0 1.2cm,clip]{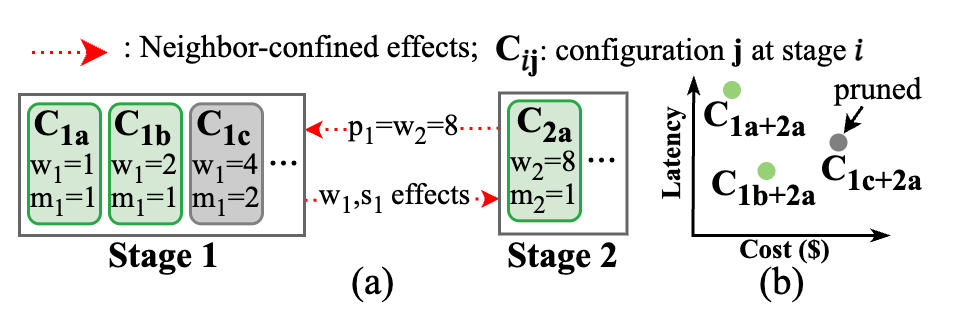}
    
    \caption{Neighbor-confined effects and local Pareto frontier.}
    \label{fig:eval-algo-example}
\end{figure}

\cref{fig:eval-algo-example} shows a detailed example of this pruning.
\cref{fig:eval-algo-example}a illustrates how the choice of worker count at stage 2 (${w_2}$) constrains feasible configurations at stage 1. Given a fixed ${w_2}$, the remaining configuration space for stage 1—including worker count, worker size, and storage type—can be enumerated and evaluated using the cost model to yield a local Pareto frontier conditioned on ${w_2}$. 
Configurations outside this frontier are pruned, as illustrated in \cref{fig:eval-algo-example}b.

    

\begin{algorithm}[t]
    \caption{Incremental Pareto Boundary Search}
    \label{alg:incremental-pruning}
    \begin{algorithmic}[1]
        \footnotesize
        \State \textbf{Input:} $\text{Card}[]$ - estimated data sizes for stages $[1..N]$
        \State \textbf{Output:} Global Pareto-optimal query execution plans
    \Procedure{IncrementalSearch}{$\text{Card}[]$}
        \State $\text{prunedSpace}[0] \gets \{\emptyset\}$
        \For{$i \gets 1$ to $N$}
            \State $\text{stageSpace} \gets \text{GenStageSpace}(\text{Card}[i])$
            \State $\text{prunedSpace}[i] \gets \emptyset$
            \For{each $\{(w_i, s_i) : \boldsymbol{m}_i\} \in \text{stageSpace}$}
                \State $\text{localSpace} \gets \emptyset$
                \For{each $\text{plan}_{i-1} \in \text{prunedSpace}[i-1]$}
                    \State $\text{plan}_{i-1}.\text{Partition} = w_i$
                    \State $\text{plan}_i \gets \text{plan}_{i-1} + \text{StageConfigs}(w_i, m_i, s_i)$
                    
                    \State $\text{localSpace}.Union(\text{plan}_i)$
                \EndFor
                \State $\text{prunedLocalSpace} \gets \mathcal{P}(\text{localSpace})$
                \State $\text{prunedSpace}[i].Union((\text{prunedLocalSpace})$
            \EndFor
        \EndFor
        \State $\text{globalParetoFrontier} \gets \mathcal{P}(\text{prunedSpace}[N])$
        \State \Return $\text{globalParetoFrontier}$
    \EndProcedure
    \end{algorithmic}
\end{algorithm}

\subsection{Cost model}
\label{sec:design:costmodel} 
In \coolName{}, \textit{Execution time} and \textit{monetary cost} of a given plan are estimated by \textit{Time Model} and \textit{Money Model}, respectively. 
The planner uses these two estimates to find the pareto frontier of plans.

\label{sec:design:timemodel} 
\subsubsection{Time model} The time model is built via a one-time offline calibration process to capture compute and storage infrastructure performance characteristics.
It has three main components: (1) operator component, (2) cloud platform component, and (3) storage service component.
This query-agnostic, compartmentalized design enables easy integration of new operators or migration across storage services/cloud environments.
For brevity, we give only a high-level overview here; full details appear in the \hyperref[sec:appendix:costmodel]{Appendix}.

\noindent
\textbf{Operator Component} calculates the time cost for each operator (e.g., scan, join, aggregation). The calculation is based on the input data cardinality, the operator's complexity, and the allocated CPU and memory resources. 

\noindent
\textbf{Cloud Platform Component} addresses the cold start challenge (\cref{sec:challenges:cost_model}) by modeling the real-world overheads of the specific serverless platform (AWS Lambda). It accounts for provider-specific behaviors, most notably cold start delays and network bandwidth patterns. 
Our empirical analysis shows that even with immediate reuse, over 10\% of workers can experience cold starts at scales of 500 or more. We model this behavior to statistically account for cold starts during cost prediction and evaluate the modeling in \cref{sec:eval:costmodel}.

\noindent
\textbf{Storage Service Component} component tackles the storage straggler challenge (\cref{sec:challenges:cost_model}) by estimating the time taken to read and write intermediate data. It factors in the volume of data, the number of concurrent I/O requests, and service-specific limitations like rate throttling. Notably, this involves measuring response latency at various levels of request concurrency to capture effects like elevated read latency and rate throttling that occur at high throughput . These measurements are then incorporated directly into the cost model and evaluated in \cref{sec:eval:costmodel}.

\subsubsection{Money model} The money model is a function that takes the execution plan and returns the cost of executing the plan. The combined total cost is the sum of the cost per stage. For a given stage, the cost of executing the stage is a sum of \textit{Worker Cost} and \textit{Storage Cost}.
\textit{Worker Cost} represents worker costs based on execution time, memory size, and invocation fees, which vary depending on the cloud provider selected. Meanwhile, \textit{Storage Cost} encompasses storage costs determined by the number of read/write operations, data volume, and the specific storage service utilized, with different storage types (e.g., object storage, file storage, block storage) carrying distinct pricing structures.

\subsection{Query Execution}
\label{sec:design:execution} 


After the SL planner selects an SL Plan using the cost model, the \textit{SL Execution Engine} executes the selected plan.
To minimize the execution time, \coolName{} employs a {\bf hybrid execution} strategy whose objective is to balance the performance benefit of compiled (as opposed to interpreted) operators used in each stage with the latency overhead of compilation.

Our hybrid execution engine starts executing the initial stages of a query using interpreted operators that are slower than compiled ones but require no compilation. While these early stages run, the code for later stages is generated, compiled, and uploaded, allowing the engine to transition to the higher-performance compiled code without stalling.

To improve performance and reduce cost, the execution engine uses several operator-level optimizations:
\squishlist
\item Scan: reads Parquet file metadata to fetch only necessary column chunks in parallel, and coalesces contiguous reads into single requests to reduce overhead~\cite{velox}. It also limits concurrent I/O per worker to mitigate S3 throttling.

\item Join: implements a partitioned hash join where data is partitioned based on the join key. To minimize write requests, partitions are combined into a single file with accompanying metadata that allows the join stage to fetch only the required partitions in parallel. 

\item Aggregation: split into local and global sub-operators. Workers first compute partial aggregates in parallel, store them to S3, these are then collected and aggregated by a global aggregation stage.
\squishend

\noindent
Additionally, \coolName{} incorporates proven techniques from prior systems, such as mitigating storage stragglers through redundant requests~\cite{lambada, starling}, and combining small writes into larger chunks to reduce requests~\cite{lambada, starling}.

\subsection{Deployment Model}
\label{sec:design:deployment} 
\coolName{}'s deployment mirrors the convenience of a Query-as-a-Service (QaaS) model, where users submit queries to run on cloud data and are billed only for the execution time.
However, \coolName{} enhances this by presenting a set of Pareto-optimal plans that illustrate the trade-offs between cost and performance. The system provides two ways to select a plan: users can either interactively choose a specific cost-performance point from a Pareto curve or pre-define a preference (such as "lowest latency"), enabling the system to automatically select and execute the most suitable option.

\section{Experimental Setup}
\label{sec:experimental-setup}

\noindent
\textbf{Platform. } We prototype \coolName{} in AWS, using AWS Lambda for compute.
For storing intermediate results between query stages, the planner consider standard S3 as well as S3 One Zone, with the choice made independently for each stage.

All experiments are run in AWS Region \texttt{us-west-2}. We note that the concepts and techniques proposed in this work are general and can be applied to any cloud FaaS platform. 

\noindent
\textbf{Workload. } Our evaluation is based on varied and representative~\cite{tpch-representative} TPC-H queries:

\noindent
\textit{Q1,6:} scan-heavy without join.

\noindent
\textit{Q4,12,14,19:} single-stage join involving different tables.

\noindent
\textit{Q5,9,16:} multi-stage join with low-cardinality aggregation.

\noindent
\textit{Q3,10,18:} multi-stage join with high-cardinality aggregation.

Prior work has shown that Q1,6,3,9,18 (all part of our evaluation) represent key performance challenges in TPC-H, and a system that performs well on them will also perform well across the entire TPC-H suite~\cite{tpch-analyzed} . 

The TPC-H queries that we exclude have elements unsupported by our code generation pipeline; this is purely an engineering effort with no anticipated effect on the overall trends. Of these queries, the majority focus on join processing and resemble Q3 and Q9. A smaller subset contain high-cardinality aggregation, similar to Q18. 


\noindent
\textbf{Dataset. } For the experiments, we use the dataset from the TPC-H benchmark suite with scale factors (SF) of 100, 1000, and 10000. The SF in TPC-H is a parameter that determines the size of the dataset in GBs. The data is stored in AWS S3 as partitioned Parquet \cite{parquet} files of each less than 1GB size in a separate directory per file along with the respective metadata. 
We use GZIP compression and plain encoding for the Parquet files. GZIP compression typically delivers good compression ratios, which helps reduce the amount of data transferred. Meanwhile, plain encoding makes the resource-intensive decompression process fast. 

\noindent
\textbf{Comparison points.} 
We are not aware of any prior serverless-native pipeline for complex analytics query processing, making direct and fair comparison with \coolName{} difficult.
Existing research either focuses exclusively on the execution engine without addressing query planning~\cite{lambada, starling}, or lacks support for complex, multi-stage queries on large datasets~\cite{astra}, or targets a radically different serverless substrate than what is offered in today's clouds~\cite{ditto}.
On the commercial side, a number of systems claim to be "serverless" but in reality run on provisioned VMs. For instance, both Snowflake\cite{snowflake-serverless} and Databricks\cite{databricks-serverless} require the user to provision a baseline set of resources and a maximum degree to which they can scale, with charges incurred as long as the provisioned resources are deployed, making these platforms unattractive for intermittent query arrival patterns that are naturally accommodated by serverless-native execution. 

The closest systems to the true serverless execution model that we target are AWS Athena~\cite{athena} and Google BigQuery~\cite{bigquery}. Both are QaaS systems that allows users to run SQL queries on data stored in, respectively, S3 and Google Cloud Storage, and run queries on-demand using pay-as-you-go billing with no resources provisioned by the user. We compare to AWS Athena, since \coolName{} also runs on AWS. 



\noindent
\textbf{Methodology.} For all experiments, we perform three runs and report the results (latency or/and cost) based on the median run in terms of latency. 
The latencies are reported as end-to-end query execution times, including query planning and actual execution time. The cost of executing the queries is based on AWS pricing at the time experiments were run.



\section{Evaluation}
\label{sec:evaluation}

\subsection{Efficacy of \coolName{} on TPC-H Q4}
\label{sec:eval:perf}

\begin{figure}[t]
    \centering
    \begin{tikzpicture}
        \node[anchor=south west, inner sep=0] (myimage) at (0,0) {
            \includegraphics[width=0.35\textwidth,trim=0cm 0.255cm 0cm 0.25cm,clip]{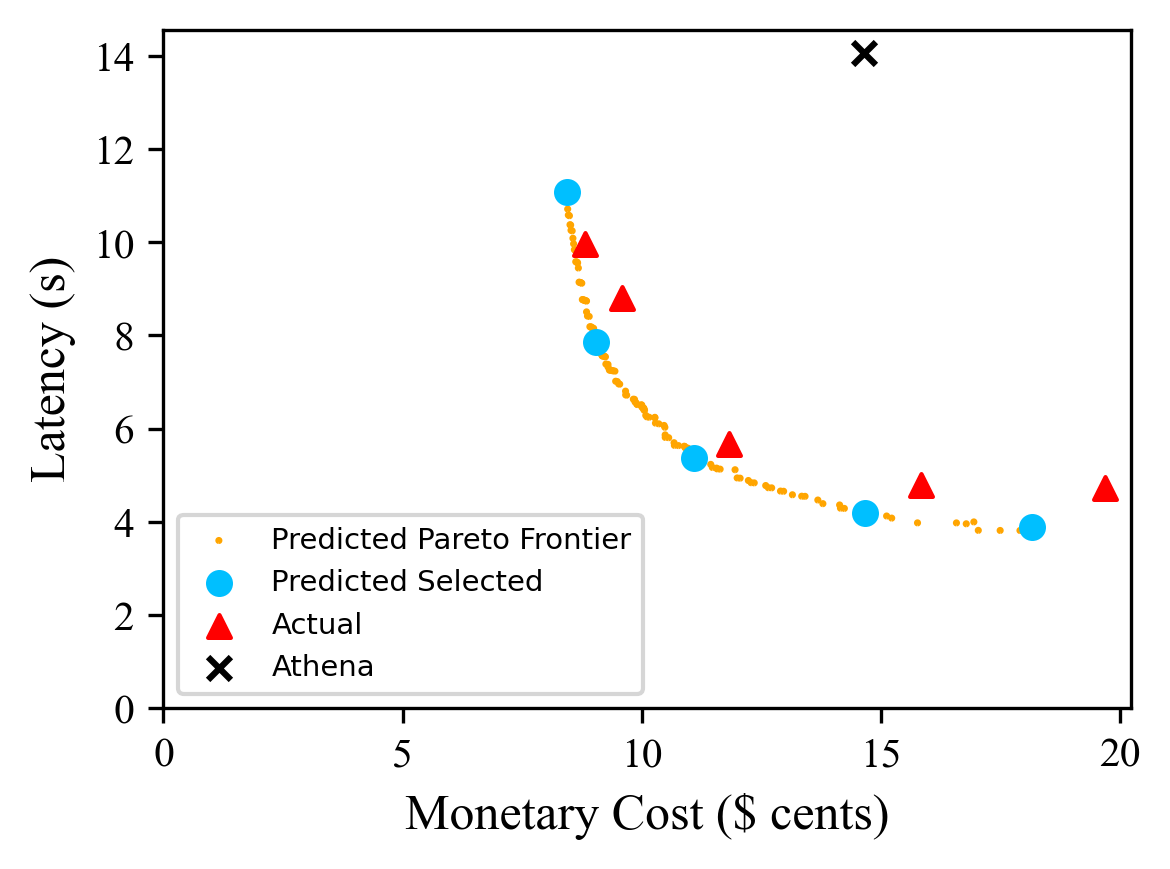}
        };
    
        \begin{scope}[x={(myimage.south east)},y={(myimage.north west)}]
            \draw[->, black, thick, -stealth] (0.7, 0.6) -- (0.63, 0.5) node[near start, right, fill=white, opacity=0, text opacity=1] {Knee Point};
        \end{scope}
    \end{tikzpicture}
    
    \caption{Pareto frontier predicted by \coolName{}, actual measurements for selected configs and Athena for Q4 at SF 1K.}
    \label{fig:big_graph}
\end{figure}

We begin our evaluation of \coolName{} with TPC-H Query 4 at SF 1K (1 TB dataset).
This multi-stage query performs filtered table scans followed by a join and aggregation.
While simple, this query is representative and straightforward to analyze.
\cref{fig:big_graph} illustrates the Pareto frontier for Q4, where the fastest configurations are also the most expensive. 
In the figure, the fastest yet most expensive configuration appears at the bottom-right, while the cheapest yet slowest lies at the top-left.
We refer to the configuration that balances cost and performance as the {\em knee-point} configuration.

\begin{figure}[t]
    \centering
    \includegraphics[width=0.48\textwidth,trim=0cm 0.33cm 0cm 0.33cm,clip]{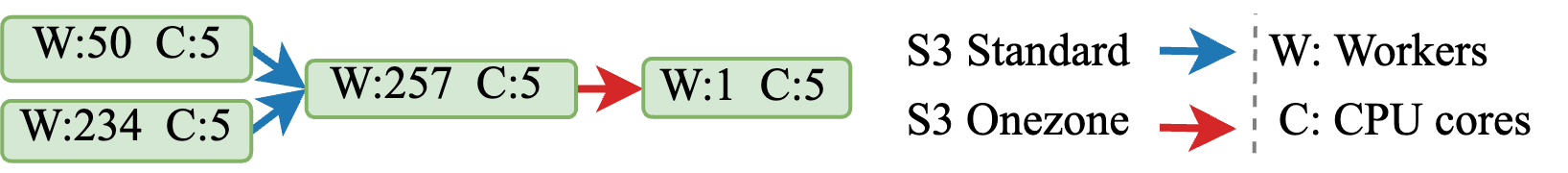}
    
    \caption{Knee-point configuration for Q4 at SF 1K.}
    \label{fig:config}
\end{figure}

\cref{fig:config} shows the knee-point configuration chosen by \coolName{}.  Notably, it uses different storage services for intermediate data exchanges--S3 Standard for the high-cardinality first stage and faster S3 OneZone for the subsequent stage.

To verify \coolName{}'s cost and latency prediction accuracy, we selected five Pareto frontier configurations (blue dots in~\cref{fig:big_graph}) and executed them on AWS Lambda using \coolName{}'s \textit{Execution Engine}.
Results validate \coolName{}'s predictions: 
Predicted and actual costs differed by less than $10\%$, and latencies by under $20\%$, demonstrating \coolName{}'s accurate estimation.



We observe that actual cost and latency are generally slightly higher than predicted due to stragglers from both workers and storage. 
On the worker side, stragglers arise from cold starts (\cref{sec:challenges:cost_model}); their incidence varies across runs, and while the cost model accounts for them probabilistically, high variance makes exact prediction difficult. 
In \cref{fig:big_graph}, all executed configurations except the {\em slowest} experience cold starts--the slowest configuration uses the fewest workers, reducing its exposure to cold-start delays.
On the storage side, stragglers primarily arise from S3 rate throttling which becomes more likely as worker counts increase. 
Both cold starts and S3 throttling are modeled probabilistically as they tend to increase execution latency and, correspondingly, cost in serverless pay-as-you-go environments.

Compared to AWS Athena, \coolName{} shows significant advantages. 
All of the evaluated configurations in \coolName{} outperform Athena in performance, with the {\em slowest} configuration on \coolName{}'s Pareto frontier being 30\% faster and 40\% cheaper than Athena. A \coolName{} configuration costing the same as Athena achieves 66\% lower latency. In fact, more than half of the Pareto-optimal configurations predicted by \coolName{} outperform Athena in {\em both} cost and performance. 

\subsection{Efficacy of \coolName{} on All Queries}
\label{sec:eval:queries}

\begin{figure}[t]
    \includegraphics[width=0.4\textwidth,trim=0 0.30cm 0 0.27cm,clip]{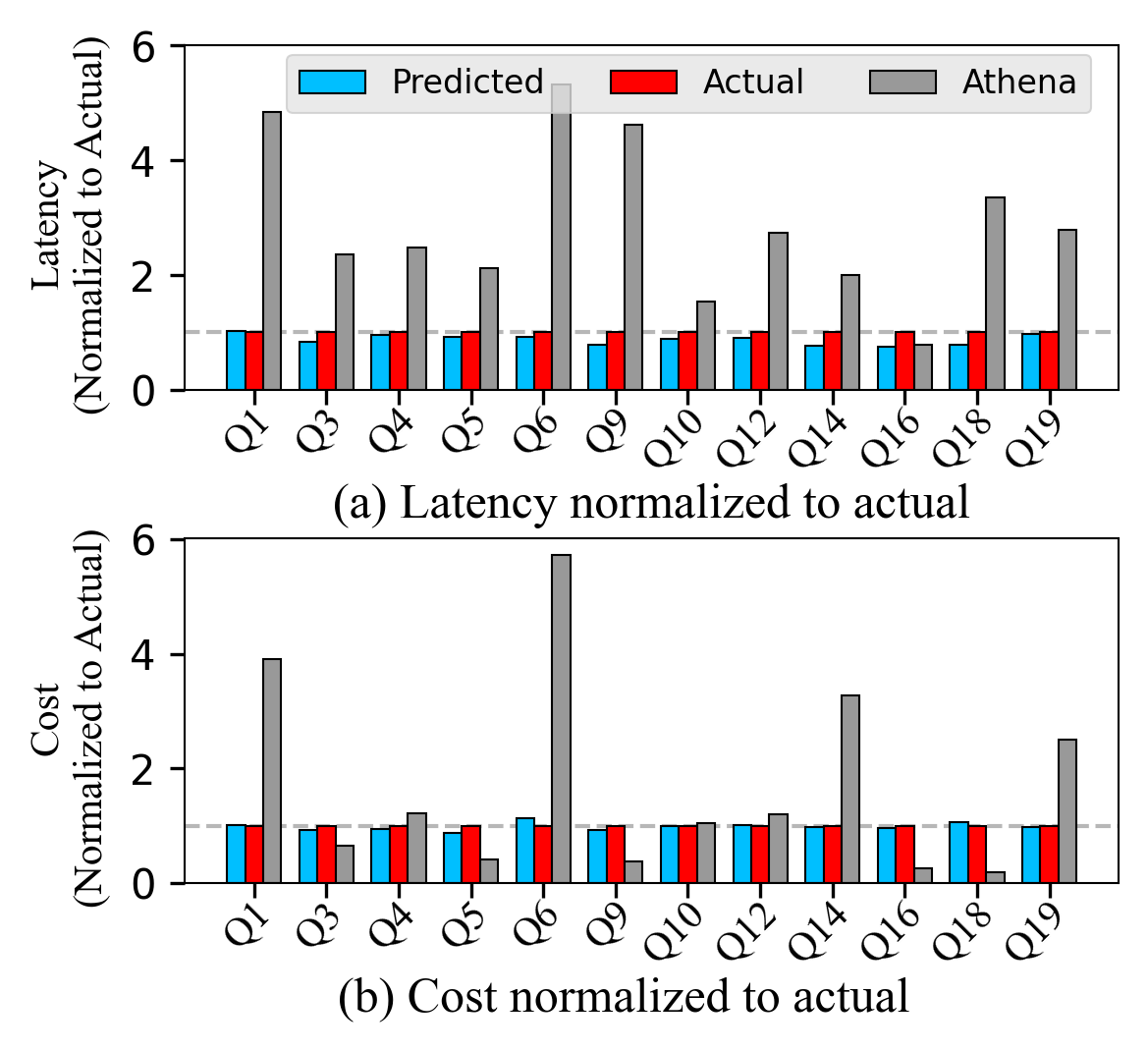}
    \caption{Predicted and actual query knee-points vs Athena}
    \label{fig:eval-knee}
\end{figure}

We extend our evaluation to all supported TPC-H queries at SF 1K.
\cref{fig:eval-knee} shows a query-by-query comparison at the predicted knee-points of \coolName{}'s Pareto frontiers.
Across all queries, predictions closely match actual measurements, with average deviations of  5\% for cost (max 13\%) and 15\% for latency (max 25\%).
Prediction accuracy is generally higher for scan and simple join queries, owing to their simpler, fewer-stage structure, versus complex join-heavy queries.
Nevertheless, \coolName{} maintains high prediction accuracy even for the most complex cases (e.g., Q9, comprising 10 stages, 5 of which are joins).

When compared to AWS Athena, \coolName{}'s knee-point configurations consistently deliver much lower latency for all but one queries (Q16, where \coolName{} is slightly slower). For cost, we find that \coolName{}'s advantage for knee-point configurations declines with increasing query complexity and exceeds Athena's cost for the most complex queries (e.g., Q3, Q5, Q9). We believe that
\coolName{}'s higher cost is due to significant inter-stage data movement in complex queries, which in \coolName{} incurs more S3 requests and longer processing times, both of which increase cost. 
In contrast, Athena's pricing model charges only for the size of scanned input data, ignoring inter-stage data movement.

\subsection{\coolName{} at Different Scale Factors}
\label{sec:eval:scales}

\begin{figure*}[t]
    \subfloat[Query 4, SF 100] {
        \includegraphics[width=0.25\textwidth,trim=0 0.25cm 0 0.22cm,clip]{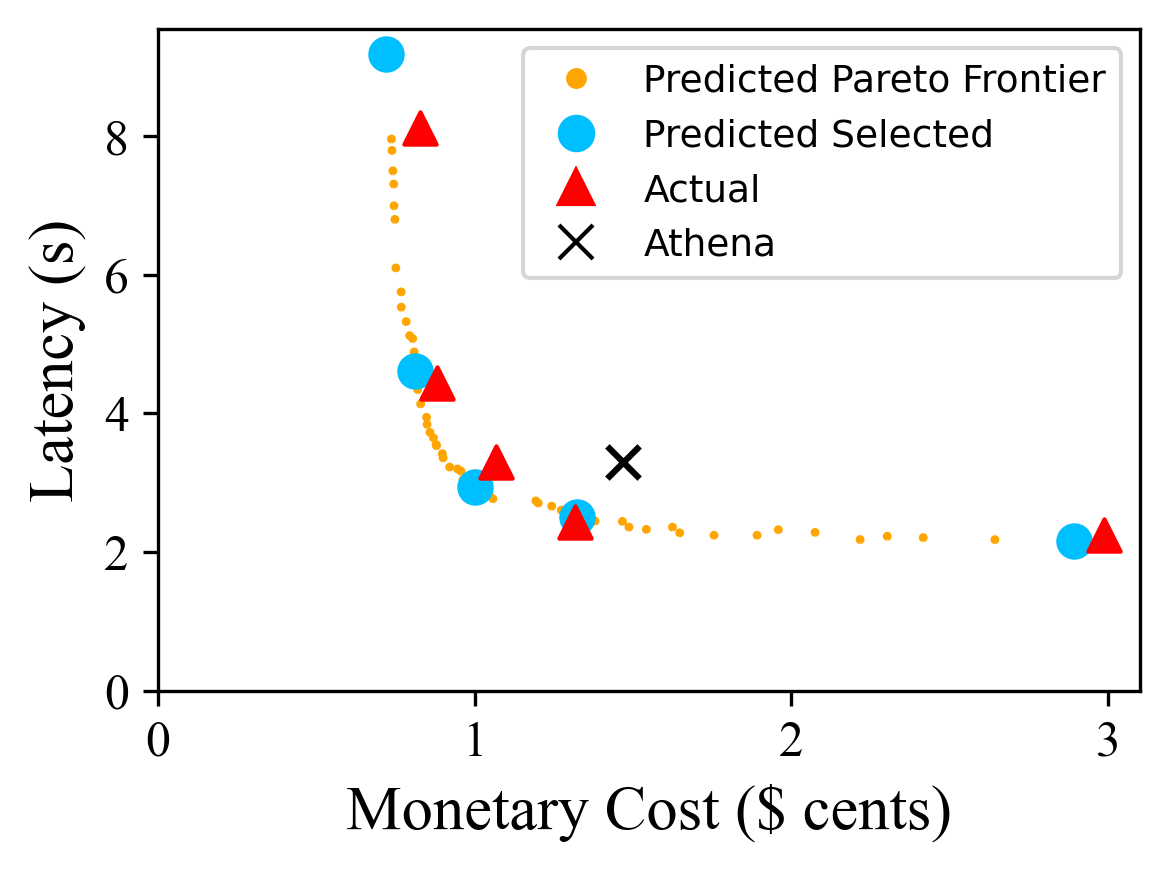}
        \label{fig:eval-q4-100}}
    \hfill
    \subfloat[Query 4, SF 10k] {
        \includegraphics[width=0.25\textwidth,trim=0 0.25cm 0 0.22cm,clip]{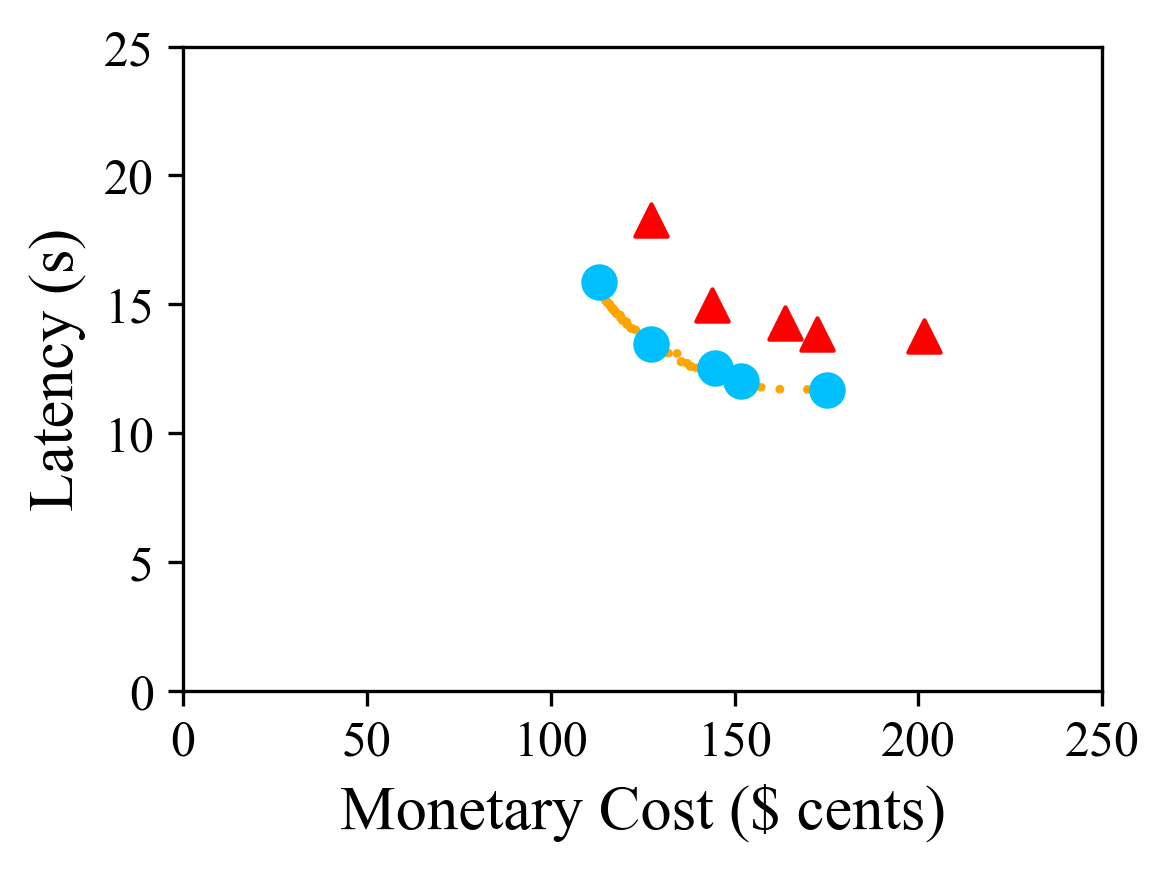}
        \label{fig:eval-q4-10000}}
    \hfill
    \subfloat[Query 14, SF 10K] {
        \includegraphics[width=0.25\textwidth,trim=0 0.25cm 0 0.22cm,clip]{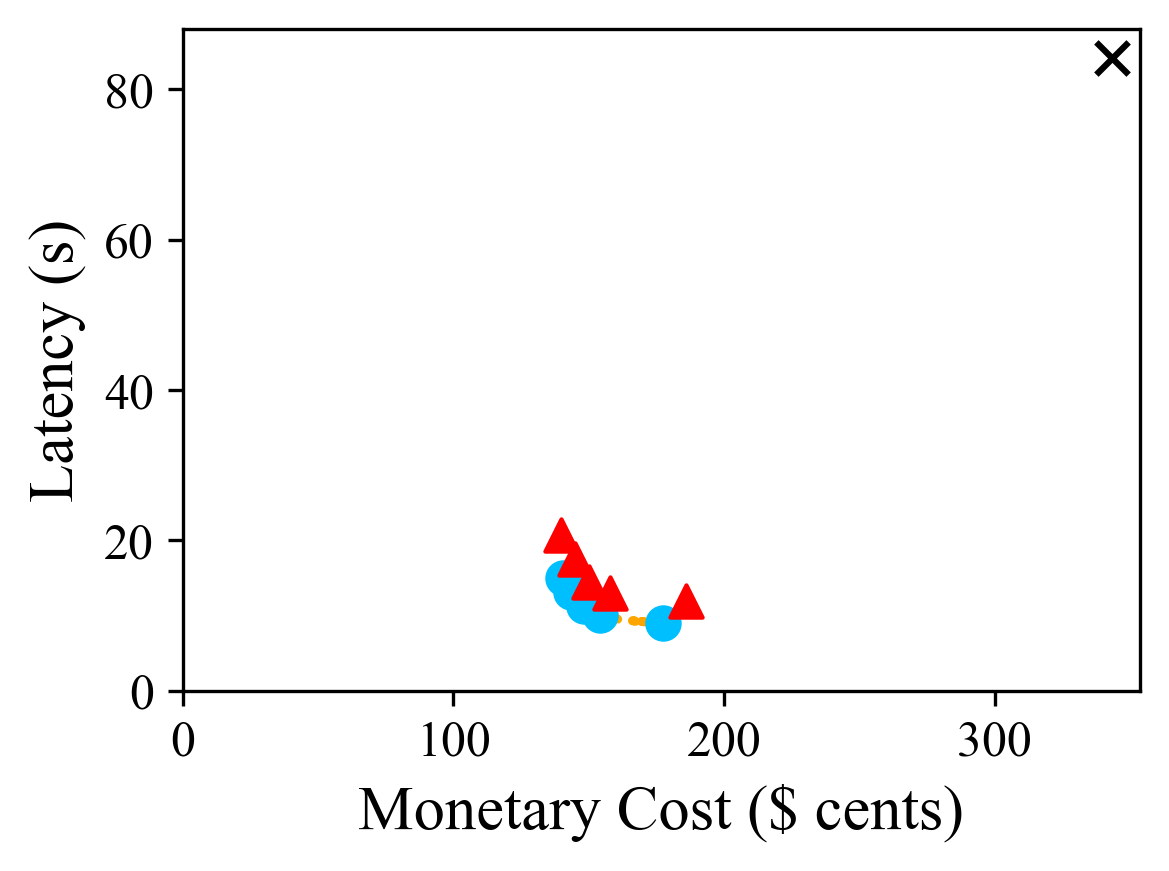}
        \label{fig:eval-q14-10000}}
    \vskip -7pt
    \caption{Pareto frontier predicted by \coolName{}, actual measurements for selected configs and Athena at SF 100 and 10K. Athena failed to complete Q4 at SF 10K. Hence, this data point is omitted, and Q14 (similar to Q4) is included instead at SF 10K.}
    
    \label{fig:eval-scales}
\end{figure*}

\begin{figure}[t]
    \includegraphics[width=0.4\textwidth,trim=0 0.3cm 0 0.36cm,clip]{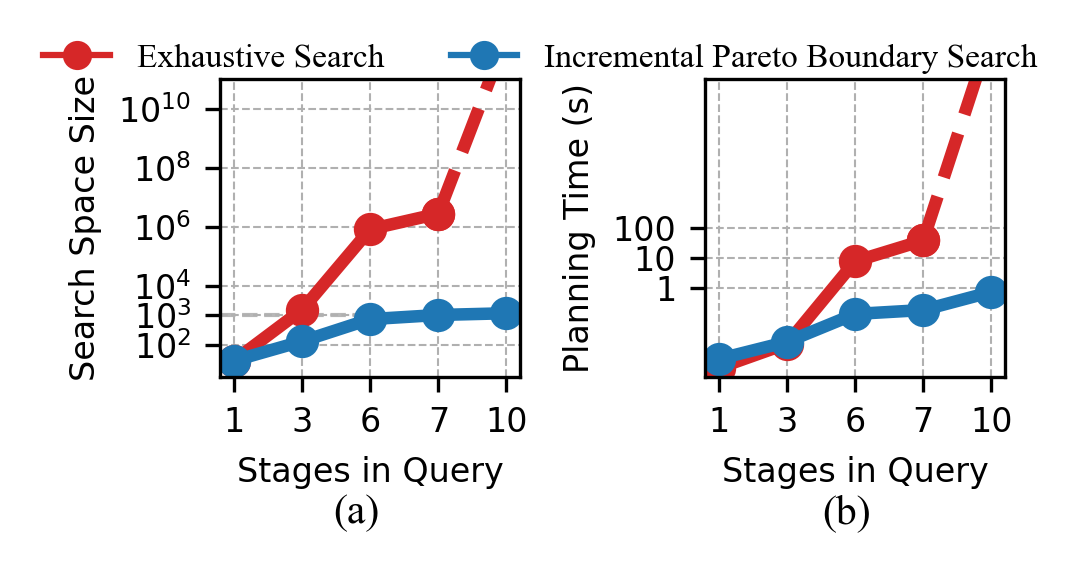}
    
    \caption{Space and time efficacy of {\em Incremental Pareto Boundary Search}. For queries with $>$7 stages, the search space size exceeds 10 billion and overflows server memory. 
    }
    
    \label{fig:eval-algo}
\end{figure}

We next study \coolName{}'s efficacy at different scale factors. We focus on join-dominated queries as they are more challenging to predict, as discussed in the previous section. 
We first focus on Q4 at SF100 (i.e., 10x smaller dataset than discussed thus far). 
As shown in \cref{fig:eval-q4-100}, \coolName{} achieves excellent accuracy in predicting the Pareto frontier. Except for the least-expensive configuration, all other configurations have very good alignment between predicted and actual latency and cost. 
For the outlier configuration, the actual latency is 12\% faster than predicted. This discrepancy arises because our cost model tries to account for potential cold start delays (\cref{sec:challenges:cost_model}). However, the smaller number of workers in this configuration did not experience any cold starts, resulting in faster execution than predicted. 

Athena performs well at the smaller scale factor. Nonetheless, \coolName{} finds four configurations with lower cost and two with better latency. One of these configurations beats Athena in {\em both} latency (by 40\%) and cost (by 12\%).

At SF 10K, \cref{fig:eval-q4-10000} shows that \coolName{} identifies a Pareto frontier, with predicted points closely matching actual measurements. At this scale, \coolName{} opts for more workers, which increases runtime variance due to cold starts and storage. As a result, deviation between predicted and actual values rises up to 13\% for latency and 15\% for cost.

Interestingly, Athena failed to complete Q4 at SF 10K despite multiple attempts.
To compare with Athena at SF 10K, we study TPC-H Q14, another join-dominated query. Athena successfully completes this one, with results in \cref{fig:eval-q14-10000}. Overall, Athena's efficiency is strikingly poor compared to \coolName{}. 
The fastest configuration of \coolName{} is 7x faster and 1.8x cheaper than Athena, while \coolName{}'s cheapest configuration is 4x faster and 2.4x cheaper than Athena. All Pareto-optimal configurations predicted by \coolName{} outperform Athena in both dimensions, demonstrating \coolName{}'s superior efficiency and cost-effectiveness for large datasets.

\subsection{\coolName{}'s Planner Efficiency}
\label{sec:eval:optimizer}

We evaluate the efficacy of ~\coolName{}'s query Planner on two fronts: (1) its ability to reduce the search space via Incremental Pareto Boundary Search, and (2) low planning latency.

\subsubsection{Search space reduction} 
\label{sec:eval:optimizer:space}

We compare search space size and planning time for all queries at SF 1000 (\cref{fig:eval-algo}) using Incremental Pareto Boundary Search (\cref{sec:design:optimizer:algorithm}) versus exhaustive search, with both approaches use heuristics from \cref{sec:design:optimizer:heuristics}.
Without the algorithm, every candidate plan must be evaluated exhaustively.
Experiments ran on an AWS EC2 c6a.8xlarge instance (32 vCPUs, 64GB RAM).
While exhaustive search requires a large server, Incremental Pareto Boundary Search is lightweight enough to run on Lambda, making its cost negligible relative to query execution.

\cref{fig:eval-algo}a shows that exhaustive search space grows exponentially with the number of stages, exceeding 10 billion configurations for 10 stages and causing out-of-memory errors.
In contrast, Incremental Pareto Boundary Search stabilizes around 1,000 configurations, independent of stage count, confirming our analysis in \cref{sec:design:optimizer:algorithm}.

We further confirmed that the Pareto-optimal configurations identified by Incremental Pareto Boundary Search are consistent with those found via the exhaustive search. Results omitted for brevity.

\subsubsection{Planning time} 
\label{sec:eval:optimizer:time}

\cref{fig:eval-algo}b shows the planning time for all queries at SF 1K (1TB dataset). \coolName{}'s planner demonstrates consistently low planning time across all queries. 
The maximum planning time of $713\text{ms}$ is observed on the most complex query, Q9.
Planning time grows linearly with the number of query stages but remains low overall, as the search space is pruned after each stage, limiting its size (see \cref{sec:design:optimizer:algorithm}).
For all evaluated queries, the planning overhead represents less than 5\% of the total execution time. 

For context, we measured Athena's planning times on the same queries and found it to be in the 100-700ms range, which is comparable to \coolName{}, thus indicating that \coolName{} achieves production grade efficiency.
\subsection{Comparison with a Query-Specific Planner}
\label{sec:eval:profile}

\begin{figure}[t]
    \includegraphics[width=0.48\textwidth,trim=0cm 0.28cm 0cm 0.24cm,clip]{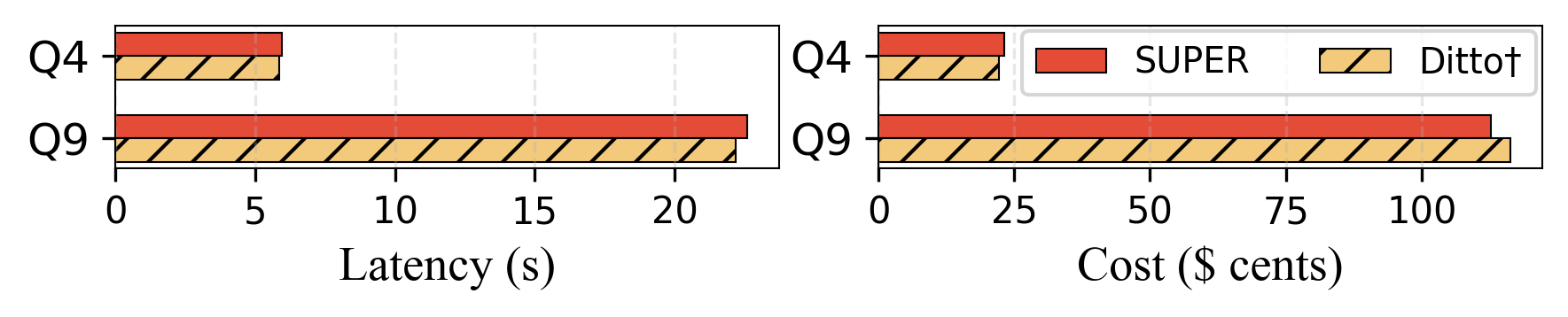}
    \caption{\coolName{} compared to Ditto at worker count chosen by \coolName{} at SF 1K. Q4: 542 workers; Q9: 1506 workers.}
    \label{fig:eval-ditto-bar}
\end{figure}

\begin{figure}[t]
        \includegraphics[width=0.45\textwidth,trim=0cm 0.27cm 0cm 0.24cm,clip]{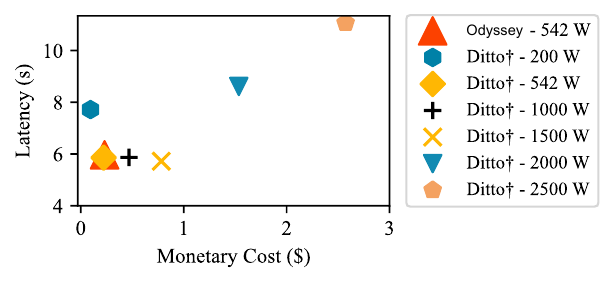}
        
    %
    
    \caption{\coolName{} compared to Ditto across worker configurations (Q4, SF 1K). Note that y-axis starts from 4 seconds.}
    \label{fig:eval-ditto-q4}
    
\end{figure}
As discussed in \cref{sec:challenges:planning}, Ditto profiles individual queries to build a query- and dataset-specific cost model.
Its query-specific planner presents a valuable comparison for \coolName{}'s query-agnostic approach, which requires no re-profiling.

In Ditto, once the query-specific cost model is built, users specify the total number of workers, which the system then partitions across stages for cost-performance efficient execution.
For comparison, we adapted Ditto to AWS Lambda, denoted as Ditto†.
Since Ditto cannot select worker sizes, we empirically found workers with 5 cores to yield the best performance and used this fixed size in our adaptation.
For fairness, we configured \coolName{} to also use 5-core workers for all stages, with S3 Standard as the storage backend.

For a given query, \coolName{}'s planner selects the total worker count (W) at the knee point, which we provide to Ditto†. Ditto† partitions this pool across stages per its model, while \coolName{} performs its own allocation. Queries are run under both configurations independently. 
\cref{fig:eval-ditto-bar} compares Q4 and Q9 at SF1K, showing similar latencies and costs, with trends representative of other queries.
Notably, despite making completely independent stage allocation decisions, \coolName{} matches Ditto†'s efficiency without query-specific profiling that Ditto relies on.

We further vary W supplied to Ditto† to evaluate the effect of total worker count (\cref{fig:eval-ditto-q4}, Q4 at SF1K). 
The point $W=542$, automatically identified by \coolName{}, yields the same cost-performance trade-off for Ditto† as for \coolName{}, confirming the optimality of \coolName{}'s knee-point selection. 
Values of W away from this optimum quickly result in rapidly diminishing cost and/or performance, highlighting the importance of selecting the correct worker count. 
Ditto†, however, cannot determine this value automatically, instead relying on the user to provide it and offering no strategy for selecting worker sizes or storage types.

In summary, \coolName{}'s query-agnostic cost model delivers the same cost-performance trade-offs as Ditto's query-specific, profile-guided approach—despite Ditto requiring \coolName{}'s worker-count input to achieve this efficiency. By eliminating per-query profiling and autonomously selecting the optimal worker count, \coolName{} requires far less planning overhead, making it more practical for diverse workloads.

\subsection{Hybrid Execution Engine Efficacy}
\label{sec:eval:hybrid-engine}

\begin{figure}[t]
    \includegraphics[width=0.43\textwidth,trim=0cm 0.28cm 0cm 0.24cm,clip]{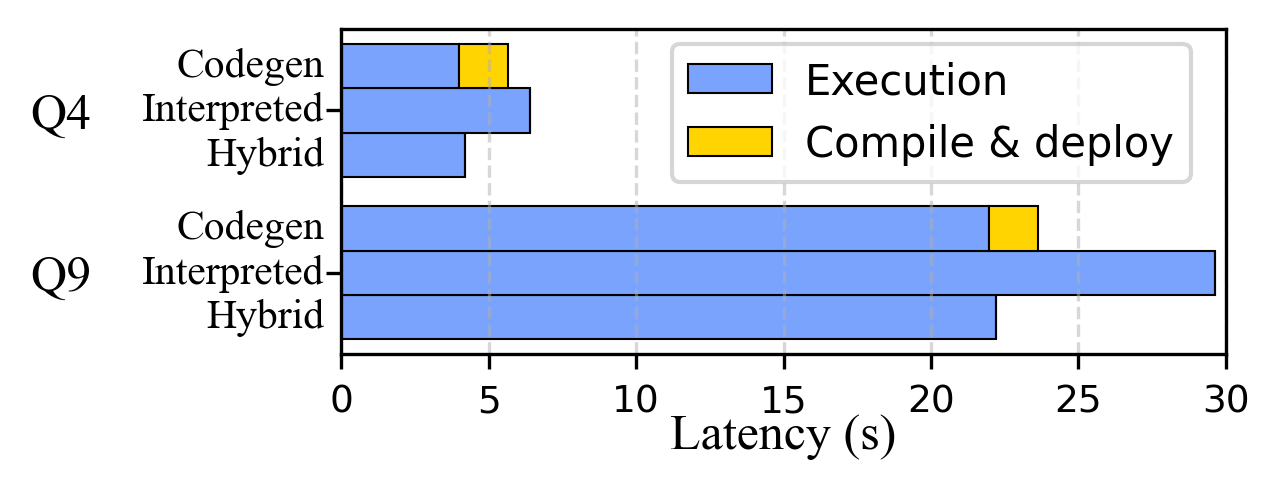}
    \caption{Query latency breakdown by execution strategy.}
    \label{fig:eval-hybrid}
\end{figure}
We next evaluate the efficacy of a hybrid query execution strategy, as discussed in \cref{sec:design:execution}.
\cref{fig:eval-hybrid} shows latency breakdown at the knee points produced by \coolName{} for Q4 and Q9 at SF 1K, using three strategies: compiled-only, interpreted-only, and hybrid.
The breakdown distinguishes execution time (actual query runtime) from compile-and-deploy time (time to compile operators and deploy to AWS Lambda).

For execution latency, compiled-only strategy is fastest, while interpreted-only is slowest, reflecting the efficiency of compiled code.
However, compile-and-deploy overhead can offset these gains, particularly for simple queries (e.g., Q4) or small datasets with short overall runtimes.
\coolName{}'s hybrid strategy mitigates this by using pre-deployed interpreted operators in the initial scan stage.
By the time scanning finishes, subsequent stages have been compiled and deployed, effectively hiding compile latency from the critical path.

While the benefit is modest for long-running queries dominated by execution time, the hybrid approach consistently avoids performance regressions and improves responsiveness for short-running queries.
By overlapping compilation with interpreted execution, \coolName{} achieves near-compiled performance without the upfront compilation cost.

\subsection{Modeling Cold Starts and S3 Throttling}
\label{sec:eval:costmodel}

\begin{figure}[t]
    \includegraphics[width=0.49\textwidth,trim=0 0.28cm 0 0.45cm,clip]{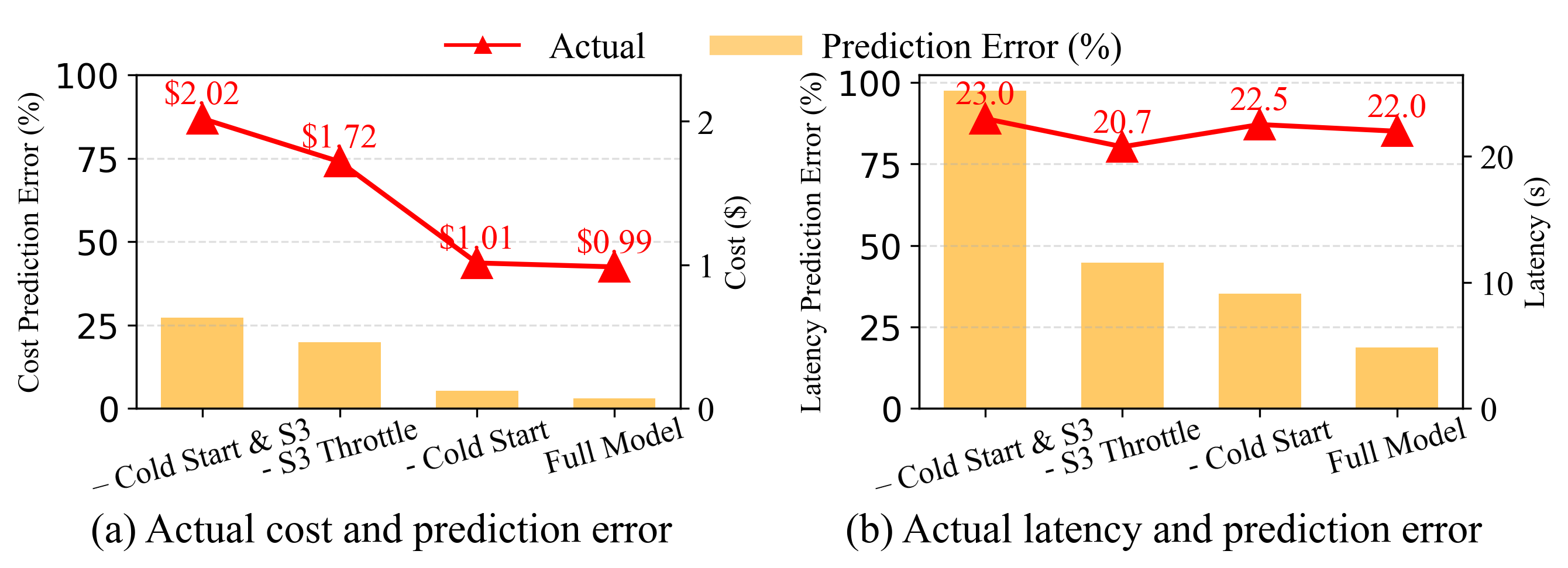}
    \caption{Actual knee-points and prediction error for Q9 at SF 1K using different cost model variants. “–” indicates exclusion of cold start and S3 throttling parameters.}
    \label{fig:eval-costmodel}
\end{figure}

Lastly, we evaluate the impact of modeling variability from cold starts and S3 throttling in our cost model.
\cref{fig:eval-costmodel} shows actual values and prediction errors on TPC-H Q9 knee points at SF 1K using four cost model variants: excluding cold starts, excluding throttling, excluding both, and the full model with all parameters.
Since model variants produce different Pareto frontiers, the selected configuration differs across models.

For cost, ignoring cold starts and S3 throttling yield configurations up to 2x more expensive (\cref{fig:eval-costmodel}a). 
The planner, unaware of these bottlenecks, over-provisions parallelism, leading to more stragglers and higher cost.
For latency, the full model achieves the lowest prediction error (<20\%), while omitting cold start and throttling modeling increases errors by up to 5x.
Cold-started workers are a key factor--e.g., Q9's full model run used 1506 workers, including 44 cold starts; ignoring these outliers underestimates latency.
Interestingly, actual latencies at predicted knee points remain similar across models, as all chosen configurations include cold-started workers.
Cost prediction errors are generally smaller than latency errors, as latency is more sensitive to stragglers, consistent with prior findings.

In summary, modeling cold starts and S3 throttling is crucial for accurate prediction and cost-efficient query planning in serverless environments.

\section{Related Work}
\label{sec:related}




\beginbsec{Distributed query engines} Distributed query engines such as PrestoDB, Trino, and SparkSQL~\cite{presto,trino,sparksql} are designed for VM clusters.
Unlike \coolName{}, their query planners are architected to efficiently exploit long-running VM clusters rather than transient serverless functions.
However, serverless environments differ significantly from VM-based clusters:they offer high elasticity and fast startup times, have limited memory per worker, and do now allow direct worker-to-worker communication. These differences necessitate a serverless-specific planner and cost model.

\beginbsec{Hybrid serverless and VM-based query engines}
Recent works have explored using serverless to supplement VM-based query processing.
For instance, Cackle~\cite{cackle} combines VMs and serverless functions: queries run on cost-effective VMs by default, but during load spikes, excess queries are offloaded to serverless workers.
Pixel \cite{pixel} uses serverless functions as a virtual accelerator for existing distributed query engines whereby individual tasks, like filters or aggregates, can be offloaded to serverless. 

\beginbsec{Data movement in serverless} Prior works have sought to address the data movement challenge in serverless. Klimovic et al. propose Pocket~\cite{pocket}, an elastic and fully managed ephemeral data store for serverless analytics.
Boxer~\cite{boxer} enables function communication via standard TCP/IP protocols and shows how this can facilitate data processing on serverless platforms. Ustiugov et al. propose XDT~\cite{xdt}, an API-preserving method for high-performance direct function-to-function data transfers. These techniques are orthogonal to our work and can be accounted for in \coolName's cost model.

\beginbsec{Cost model for serverless analytics} 
Astra~\cite{astra} proposes a cost-model-based approach for advising serverless configurations for map-reduce jobs.
While effective for simple jobs (e.g., WordCount or Sort) with one mapper, one or more reducers, and small datasets, it does not scale to complex SQL queries.
In contrast, \coolName{} targets multi-stage query pipelines on datasets up to 10TB, providing a serverless-native planner and cost model, along with a robust execution engine that handles both worker and storage stragglers.


\section{Conclusion}
\label{sec:conclusion}

Serverless data analytics present new challenges but also novel degrees of freedom compared to traditional analytics execution environments, resulting in a complex optimization space with vast cost and performance implications. This work introduces \coolName{}, a serverless-native end-to-end data analytics pipeline that integrates a query planner, cost model and execution engine. \coolName{} is able to map the vast space of possible plans for a given query, rapidly identify the Pareto-optimal plans using its planner and cost model, and efficiently execute the chosen plan via its execution engine. 

\bibliographystyle{ACM-Reference-Format}
\bibliography{bib.bib}
\clearpage
\section*{Appendix}
\label{sec:appendix}
\subsection*{Detailed cost model}
\label{sec:appendix:costmodel}

Section 5.3 provided a high-level introduction to the cost model utilized by \coolName, outlining its role in estimating execution time and monetary cost to guide the optimizer in selecting Pareto-optimal query plans. This Appendix provides a comprehensive, detailed formulation which was omitted from the main text due to space constraints. 

As discussed in Section 5.3.1, there are three components of the cost model: (1) the operator component (2) the compute platform component (3) the storage service component. Together, these components enable \coolName to accurately estimate execution time and monetary costs across different configurations. To distinguish the terms corresponding to each component, we color code them using following colors: 
\begin{itemize}
    \item \textcolor{cyan}{Cyan for the operator component}
    \item \textcolor{purple}{Purple for the compute platform component}
    \item \textcolor{teal}{Teal for the storage service component}.
\end{itemize}

The cost model computes execution time at a worker granularity. To estimate the total query execution time, the complete query plan is traversed, combining the execution times of individual workers based on their dependencies and parallelism to generate an accurate end-to-end execution time prediction.

\subsection*{Time model}
\label{sec:appendix:timemodel}
The time model is a function that takes the execution plan and returns the execution time. For a given stage, the execution time taken by the worker is given by the following equation:
{
\begin{equation}
    t_{worker} = t_{inv} + t_{fetch\_process} + t_{output}
\end{equation}
}

All terms in this equation, as well as in subsequent equations, are defined in detail in Tables ~\ref{tab:time_terms} \& ~\ref{tab:add_terms}.

\textbf{Invocation. } The time taken by the driver to invoke the workers is given by the following equation:
{
\begin{equation}
    t_{inv} = clientInvDelay(W) + \mathcolor{purple}{providerInvDelay(W)}
\end{equation}
}


The time taken by the client and provider to invoke the workers is given by the following equations:
\setlength{\abovedisplayskip}{5pt}  
\setlength{\belowdisplayskip}{0pt}  
{
\begin{equation}
    clientInvDelay(W) = \frac{W}{clientInvRate}
\end{equation}
}
{
\begin{equation}
    \mathcolor{purple}{providerInvDelay(W) = 40 + ReLU(W-1000)*10 ms}
\end{equation}
}

Here, $ReLU(x)=max(0, x)$. When the number of workers is less than the default \textit{total concurrency limit} of the 1000~\cite{lambda-scale-default}, 40ms is the time taken by the provider to invoke the workers. Once the limit is surpassed, the provider allocates 1000 workers every 10 seconds, i.e., 10ms per worker~\cite{lambda-scale-limit}.  


\textbf{Read + Process data. } The allocated data to a worker is split into chunks. The fetch data and process data tasks are interleaved. Hence, the time depends on the slower of the two tasks. If fetch takes longer than the process, the process time is hidden and vice versa.
\begin{equation}
    t_{fetch\_process} = max(t_{fetch}, t_{process})
\end{equation}
$t_{fetch}$ depends on the latency (time to the first byte) of the external storage and the bandwidth of the worker. As per our measurements, AWS Lambda workers can fetch the data from storage at 300 MB/s for the first 150MB and 70 MB/s for the subsequent data.
\begin {equation}
    t_{fetch} = \mathcolor{teal}{Lat_{storage}} + \mathcolor{purple}{\begin{cases} \frac{150}{300} + \frac{d_{input}-150}{70} & \text{if } d_{input} > 150
    \\ \frac{d_{input}}{300} & \text{if }  d_{input} <=150 \end{cases}}
\end{equation}

\begin {equation}
    t_{process} = t_{decompress} + \mathcolor{cyan}{t_{process\_op} }
\end{equation}

$t_{decompress}$ depends on the algorithm used, compression ratio, size of the data, and the number of cores available. $t_{process\_op}$ depends on the operator being executed, the size of the data, and the number of cores available.

\textbf{Output. } The time taken by the workers to compress and store the output to storage.
\begin{equation}
    t_{output} = t_{compress} + t_{store}
\end{equation}


$t_{compress}$ depends on the algorithm used, compression ratio, size of the data, and the number of cores available. $t_{store}$ is identical to $t_{fetch}$ with $d_{output}$ being the size of the output data.

\textbf{Storage latency:} Storage latency ($\mathcolor{teal}{Lat_{storage}}$) is a component in our cost model that accounts for the time it takes to initiate data transfer from external storage services. This latency varies based on several factors including request concurrency, data size, and service-specific throttling policies.

\begin{equation}
    \mathcolor{teal}{Lat_{storage} = baseLatency + throttledLatency}
\end{equation}
Where $\mathcolor{teal}{baseLatency}$ represents the minimum time required to initiate a data transfer from the storage service under normal conditions, and $\mathcolor{teal}{throttledLatency}$ accounts for additional delays that occur only when the service experiences high request volumes (above 5500 according to our experiments). Throttled latency is given by:

\begin{equation} 
\small
\mathcolor{teal}{
    \text{LatThrottled} =
\begin{cases}
  a \times e^{b \times \left(\frac{\text{TotalReqPerSec}}{5500} - 1\right)} & \text{if } \text{TotalReqPerSec} > 5500 \\
  0 & \text{otherwise}
\end{cases}}
\end{equation}
Where $\mathcolor{teal}{a}$ and $\mathcolor{teal}{b}$ are constants determined through empirical measurements of the storage service's throttling behavior. In our experiments with AWS S3, we found $\mathcolor{teal}{a = 0.65}$ and $\mathcolor{teal}{b = 0.66}$ provide accurate predictions of throttling-induced latency.

$\text{TotalReqPerSec}$ represents the total number of requests per second made to the storage service during the execution of stage.
This model captures the non-linear increase in latency that occurs when storage services apply throttling under high request rates, which is a critical factor in accurately predicting execution times for data-intensive serverless analytics workloads.

\begin{table}[t]
\centering
{ 
\begin{tabular}{|l|p{5.7cm}|}
\hline
\textbf{Term}                        & \textbf{Time Taken for}                                                                                                                                                  \\ \hline
$ t_{worker} $                       &  Entire process by the worker.                                                                                                            \\ \hline
$ \mathcolor{purple}{t_{inv}} $                          &  Driver to invoke the workers.                                                                                                                  \\ \hline
$ t_{fetch\_process} $        & Maximum duration to fetch or process data.                                                                                                        \\ \hline
$ t_{output} $                       &  Workers to compress/store data to external storage.                                                                                 \\ \hline
\scriptsize $ clientInvDelay(W) $                &  Client to invoke $W$ workers.                                                                                                                  \\ \hline
\scriptsize $ \mathcolor{purple}{provInvDelay(W)} $              &  Provider to invoke $W$ workers.                                                                                                                \\ \hline
$ t_{fetch} $                        &  Workers to fetch the data.                                                                                                            \\ \hline
$ \mathcolor{cyan}{t_{process}} $                      &  Workers to process the data, including decompression \& processing operations.                                                       \\ \hline
$ t_{decompress} $                   &  Workers to decompress  data.                                                                                                                \\ \hline
$ t_{process\_op} $                  &  Workers to perform the operation on the data.                                                                                        \\ \hline
$ t_{compress} $                     &  Workers to compress the output data before storing.                                                                                             \\ \hline
$ t_{store} $                        &  Workers to store data to external storage.                                                                         \\ \hline
\end{tabular}
}
\caption{Definitions of time consumption terms}
\label{tab:time_terms}
\end{table}

\begin{table}[t]
\centering
{ 
\begin{tabular}{|l|p{5.7cm}|} 
\hline
\textbf{Term}                        & \textbf{Definition}                                                                                                                                                  \\ \hline
$ W $                                & Number of workers.                                                                                                                                               \\ \hline
\scriptsize $ clientInvRate $      & The rate at which the client can invoke the workers.                                                                                                             \\ \hline
$ N_{reqs} $                         & Number of read requests made by the workers.                                                                                                                     \\ \hline
$ \mathcolor{teal}{Lat_{storage}} $                         & Latency of the external storage.                                                                                                                      \\ \hline
$ d_{input} $                        & Size of the input data to be fetched by the workers.                                                                                                              \\ \hline
\end{tabular}
}
\caption{Definitions of additional terms} 
\label{tab:add_terms}
\end{table}

\subsection*{Monetary model} 
The money model is a function that takes the execution plan and returns the cost of executing the plan. The combined total cost is the sum of the cost per stage. For a given stage, the cost of executing the stage is given by the following equation:
\setlength{\abovedisplayskip}{5pt}  
\setlength{\belowdisplayskip}{5pt}
\begin{equation}
    c_{stage} = c_{workers} + c_{storage}
\end{equation}
Where $c_{worker}$ is the cost of the workers and $c_{storage}$ is the cost of the cloud storage. The cost of workers is given by the below: 
\setlength{\abovedisplayskip}{5pt}  
\setlength{\belowdisplayskip}{5pt}
\begin{equation}
\small
    c_{worker} = \sum \limits_{i=1}^{W} \mathcolor{purple}{c_{worker\_inv} + c_{worker\_duration}} * t_{worker} * Size_{worker}
\end{equation}
Where $\mathcolor{purple}{c_{worker\_inv}}$ is the cost of invoking a worker, $\mathcolor{purple}{c_{worker\_duration}}$ is the cost of a worker per second per GB of memory, $t_{worker}$ is the time taken by the worker to execute the stage, $Size_{worker}$ is the memory allocated to the worker and $W$ is the number of workers. The cost of storage is given by the following equation:
\begin{equation}
\small
    c_{storage} = \sum \limits_{i=1}^{N_r} \mathcolor{teal}{c_{storage\_req}} + 
    \sum \limits_{i=1}^{N_w} \mathcolor{teal}{c_{storage\_req}} + d_{output} * \mathcolor{teal}{c_{storage\_data\_write}}   
\end{equation}
Where $\mathcolor{teal}{c_{storage\_req}}$ is the cost of a request to storage, $\mathcolor{teal}{c_{storage\_data\_write}}$ is the cost GB of data written to storage, $d_{output}$ is the size of the output data in GB and $N_w$ and $N_r$ is the number of write and read requests made to storage respectively.
\end{document}